# Single-molecule time-resolved spectroscopy in a tunable STM nanocavity

Jiří Doležal[1#*], Amandeep Sagwal[1,2], Rodrigo Cezar de Campos Ferreira[1], Martin Švec[1,3*]

[1] Institute of Physics, Czech Academy of Sciences; Cukrovarnická 10/112, CZ16200 Praha 6, Czech Republic

[2] Faculty of Mathematics and Physics, Charles University; Ke Karlovu 3, CZ12116 Praha 2, Czech Republic

[3] Institute of Organic Chemistry and Biochemistry, Czech Academy of Sciences; Flemingovo náměstí 542/2. CZ16000 Praha 6, Czech Republic

[#] Present address: Institute of Physics, École Polytechnique Fédérale de Lausanne, CH-1015 Lausanne, Switzerland

*Corresponding authors: dolezalj@fzu.cz, svec@fzu.cz

Abstract:

The spontaneous fluorescence rates of single-molecule emitters are typically on the order of nanoseconds. However coupling them with plasmonic nanostructures can substantially increase their fluorescence yields. The confinement between the tip and sample of a scanning tunneling microscope creates a tunable nanocavity, an ideal platform for exploring the yields and excitation decay rates of single-molecule emitters depending on the coupling strength to the nanocavity. With this setup we estimate the excitation lifetimes from the direct time-resolved measurements of the fluorescence decays of phthalocyanine adsorbates, decoupled from the metal substrates by ultrathin NaCl layers. It is found that nanosecond-range lifetimes prevail for the emitters away from the nanocavity, whereas for the tip approached to a molecule, we find a substantial effect of the nanocavity coupling, which reduces the lifetimes to a few picoseconds. An analysis is performed to investigate the crossover between the far-field and tip-enhanced photoluminescence regimes. This approach overcomes the drawbacks associated with the estimation of lifetimes for single molecules from their respective emission linewidths.

Coupling single-molecule emitters with the electromagnetic field of an optical cavity can stimulate their excitation and radiative decay rates, increasing the fluorescence yields and bringing them closer to the applications in the GHz range. Such enhancement is often realized in the vicinity of plasmonic structures of subwavelength size such as sharp metal tips,[1] colloids[2] or bowtie nanoantennas,[3] where the electric field is strongly localized and intensified. In the regime of weak coupling between the emitter and the cavity, the Purcell effect[4] will cause shortening of the radiative lifetime, broadening, and redshift of the emission line (known as the Lamb shift).[5] On the other hand, in the limit of strong coupling, energy is coherently exchanged between the molecule and the plasmon. This has also been demonstrated in nanoparticle-on-mirror[6] experiments, by detecting characteristic Rabi oscillations and the emission line-splitting.[7] However, it remains a challenge to achieve control over the coupling, exact geometry of the molecule in the nanocavity and the resulting behavior. The scanning tunneling microscope (STM) provides an ideal platform where positioning of the molecules within the nanocavity formed by the tip and a substrate can be performed with outstanding precision. STM combined with an optical detection scheme represents a prospective methodology that can provide benchmarks of the coupling strength and fundamental insights in the area of the light-matter interaction.

First experimental attempts to measure the radiative lifetime from the electroluminescence of single molecules in STM nanocavity utilized a second-order photon correlation technique, the Hanburry-Brown-Twiss interferometry, which confirmed the expected single-photon character of the emission and found an exponential decay half-life on the order of hundreds of picoseconds.[8,9] Similar results have been obtained with the RF-phase fluorometry.[10] However, in these experimental setups, it was not possible to distinguish the exciton decay rates from the dynamics of charge carrier transport preceding the formation of the excitons.[11,12] In other reports, excitation lifetime values were estimated from the emission linewidths, setting them in the range of hundreds of femtoseconds.[13,14] Also this approach has limitations, stemming from possible exciton-phonon coupling occurring in chirally adsorbed molecules and instrument resolution limits, which generally lead to a secondary apparent peak broadening, obscuring the intrinsic effect of the lifetime.[15,16] Possibly because of that, so far only the lifetimes for the symmetrically-adsorbed free-base phthalocyanine in the STM nanocavity have been reliably determined, using peak widths in resonant photoluminescence spectroscopy.[15,16]

Therefore, a direct method to excite an emitter and measure its decay rate as a function of the nanocavity geometry would be more adequate, in order to exclude the charge carrier dynamics, spectral line broadening effects, and to distinguish the role of the far-field contributions to the signal[17]. We use the STM with optical excitation and detection capabilities to perform direct time-correlated single photon counting (TCSPC) using photoluminescence of single zinc- and magnesium-phthalocyanines (ZnPc, MgPc). With this technique, we could determine the upper limit of the excitation lifetimes of the molecule in the tunable nanocavity and investigate the transition region between the micro-photoluminescence (µPL) and tip-enhanced photoluminescence (TEPL), showing a dramatic effect of the electromagnetic field confinement on the lifetime.

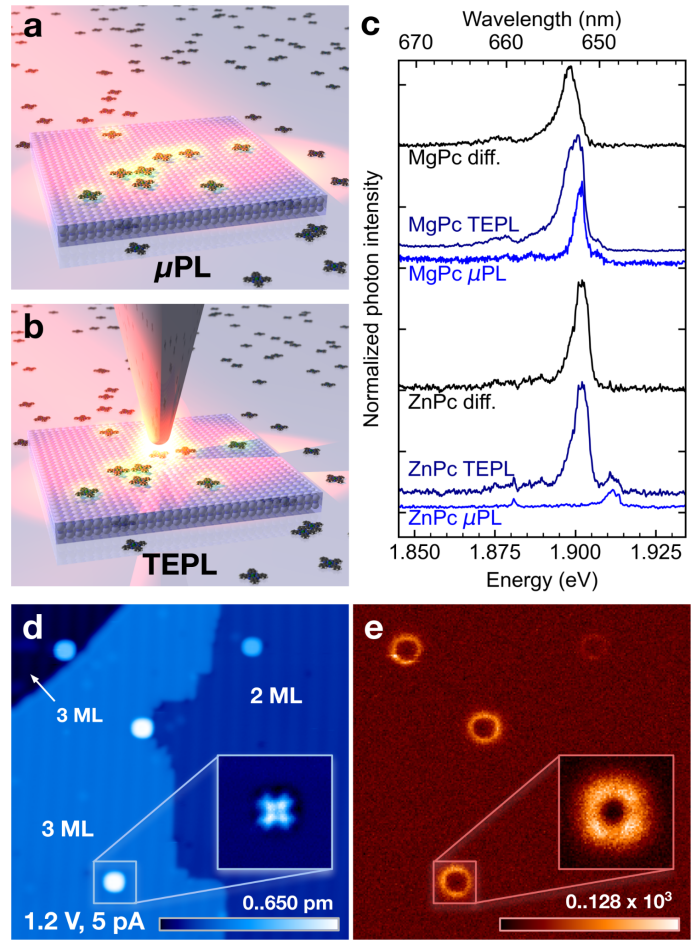

*Fig.1: Schemes of a) the far-field µPL experiment with molecules on a decoupling layer and b) the near-field + far-field TEPL experiment of a single molecule. c) The comparison of the µPL, TEPL and difference spectra for the Mg- and ZnPc adsorbed on 2-3ML NaCl/Ag(111). The MgPc TEPL spectrum was measured at 1.2 V and 1.7 pA setpoint and the µPL spectrum was obtained with the tip retracted 1 nm from the setpoint. The ZnPc TEPL was measured at 1.2 V and 5 pA setpoint and the µPL spectrum was obtained with the tip retracted 120 nm from the setpoint. d) STM topography of MgPc/2-3 ML NaCl/Ag(111), size 44 x 44 nm$^2$, tunneling conditions 1.2 V, 5 pA, 3 ms per pixel e) APD map of the identical area, taken simultaneously and rebinned from 512 x 512 to 256 x 256 pixels for better contrast. The insets in d,e) are taken in the constant-height mode in the transport gap at -0.7 V, maximum current is 1 pA, total laser power 100 µW, resulting in peak rate 130 x 10$^3$ events/point, image size 4 x 4 nm$^2$, the original APD map was re-binned from 128 x 128 to 64 x 64 pixels.*

For the photoluminescence experiments, we have prepared systems with ultralow concentrations of MgPc and ZnPc species adsorbed on 2-3 ML NaCl/Ag(111) and on clean Ag(111). The morphology of the as-prepared NaCl/Ag(111) substrate consists of three different major landscapes, *i.e.* the 3 ML, 2 ML-NaCl islands and the bare Ag(111).[11,14,15,18–20] In the µPL (far-field) experiments performed by irradiating these samples directly with a focused laser beam of 1.96 eV energy photons (schematically depicted in Fig.1a), we observe fluorescence signals only from the systems with NaCl decoupling layers, whereas the molecules on bare Ag(111) do not contribute, because of nonradiative quenching of the excitations to the substrate.[21] The MgPc signature in µPL appears as a single emission line at 1.90 eV with 4 meV full width at half maximum (see Fig.1c) and can be attributed to the transition from the excited state to the ground state. In the case of ZnPc, a peak with a similar linewidth is observed at 1.91 eV and another much narrower peak at a lower energy of 1.88 eV appears. The former peak broadening can be attributed to the exciton-libron coupling as described in our previous study.[15] In contrast, the peak at 1.88 eV is very narrow and might originate from molecules in symmetrical configurations, *e.g.,* on step-edges, $Na^+$ ions or pinned to defects. Since the estimated number of molecules illuminated by the focused laser light is between $10^4$-$10^5$, one would also expect the influence of inhomogeneous broadening[22] stemming from nonequivalent adsorption configuration of each molecule in the ensemble due to incommensurability of NaCl overlayer with the Ag(111) and local variations in the electronic properties of the substrate. Our energy resolution puts its energy range below 0.3 meV in agreement with previous observations.[22,23]

Upon closing the distance between the STM tip apex and a molecule well below 1 nm, new plasmonic modes are created, corresponding to the gap nanocavity between the Ag tip and the surface. They mediate an efficient bidirectional coupling between the electromagnetic near-field of the nanocavity coupled to the molecular emitter and the far-field of the irradiation and detection beams, thus permitting excitation and detection of single-molecule photoluminescence for the species that are electronically detached from the metal substrate (see Fig.1b). These conditions define the TEPL mode, which is however also inherently including the *μ*PL contributions due to the diffraction-limited size of the focused laser spot. With the tip in the vicinity of the MgPc or ZnPc molecules adsorbed on 3 ML NaCl, characteristic photoluminescence peaks emerge near the 1.9 eV mark. These relatively broad peaks are apparently superimposed over their corresponding *μ*PL backgrounds that can be subtracted in order to get the single-molecule contributions originating in their respective $S_1$-$S_0$ transitions, as described previously.[24] The individual molecule emission peaks are redshifted for both Mg- and ZnPc (by 10 and 3 meV, respectively) and also broadened compared to the *μ*PL measurements, in agreement with the reports of Yang et al.[11]

The spatial distribution of the TEPL yield can be visualized since the *μ*PL background signal is independent of the lateral nanocavity position over the sample. We record the photon rate by an APD filtered to a spectral range of 642-662 nm and scan the tip over the NaCl-covered sample area, first using the STM topography mode. The STM image presented in Fig.1d shows a step between 3 ML and 2 ML of NaCl/Ag(111) with a few MgPc molecules distributed in the scanned area. The APD photon map in Fig.1e shows the PL of the molecules as hollow circular features, corresponding to a single molecule each. These features are well-pronounced at 3 ML NaCl, but at the limit of detection at 2 ML due to a significantly higher degree of the electronic decoupling of the molecules from the substrate. A detailed constant-height mapping with the APD (the Fig.1e inset) reveals a characteristic donut-shaped pattern of the emission with hints of four lobes. This appearance has been observed in the previous work and explained theoretically by variation of coupling between the emitter electronic transition density and the nanocavity.[11,25] We have not observed any TEPL signal from the molecules adsorbed directly on Ag(111) due to the non-radiative quenching of their excited states by the substrate.[21] It has been reported that at a very close tip-sample distances, in particular when the molecule is contacted or even lifted up with the tip, the Raman signal[26,27,28,29] can be dramatically enhanced on NaCl or bare metal substrates. However, these conditions have been deliberately avoided in our study.

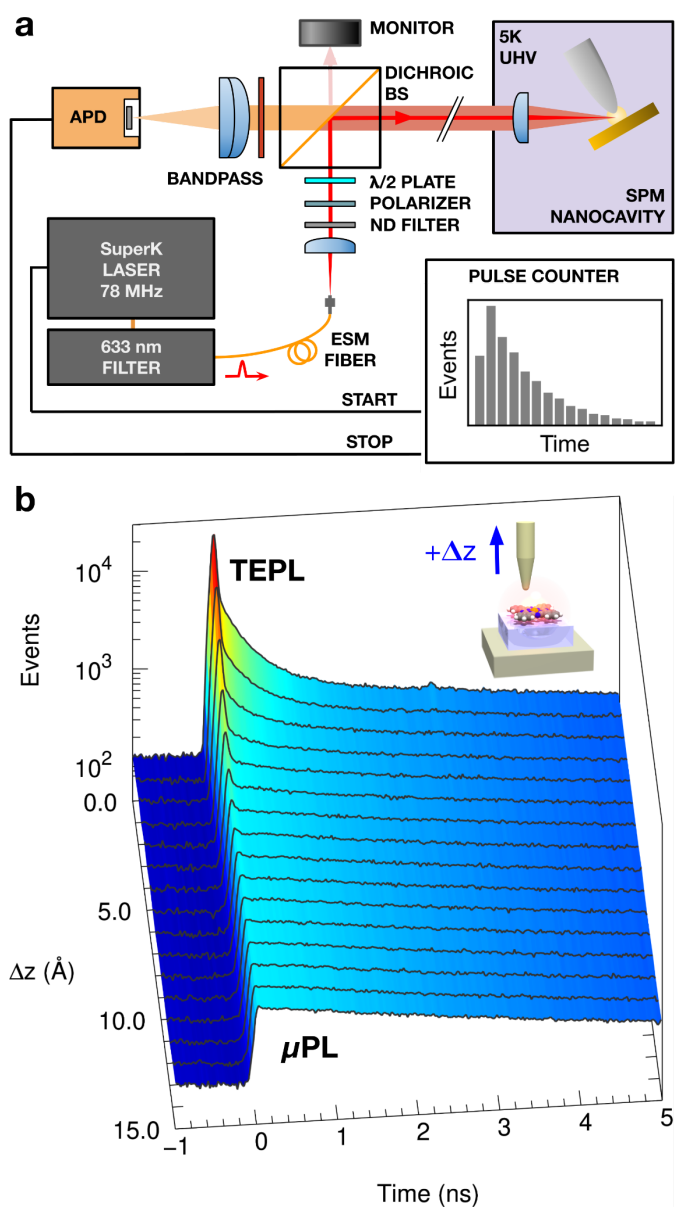

*Fig.2: a) Scheme of the TCSPC setup for the measurement of the transition between µPL and TEPL regimes with a tunable nanocavity, with a pulsed laser and an avalanche photon detector (APD). b) The MgPc $S_1$-$S_0$ photoluminescence emission photon-arrival time histograms as a function of the tip-sample distance, crossing from the µPL to the TEPL regime. The Δz = 0 setpoint corresponds to tunneling conditions of 1.2 V, 60 pA on NaCl 1.4 nm away from the molecular center. The total laser power was 100 µW.*

We employ the APD and the pulsed laser to perform the TCSPC of photoluminescence on the phthalocyanines in both TEPL and µPL regimes, as shown in the scheme in Fig.2a. High pulse repetition rates of the laser allow us to obtain good statistics of the photon-arrival delays, recorded and binned to histograms using a fast pulse counter connected to the APD and synchronized with the laser seed pulse. The spectral region of the APD measurement was filtered to the 649-673 nm region of the Mg/ZnPc main emission lines. The dependence of the MgPc decay rates on the vertical tip-sample displacement (*Δz*) in Fig.2b reveals a dramatic change of the character of the time traces. At large tip-sample distances in the µPL regime, they have a single-step shape indicative of a surprisingly long exponential decay, whereas with the tip in the vicinity of the molecule in the TEPL regime, a very fast decay appears on top at the step onset. Such a rise in intensity is a clear indication that the photoluminescence enhancement effect takes place due to the intensification of the electromagnetic field within the narrowing nanocavity gap.

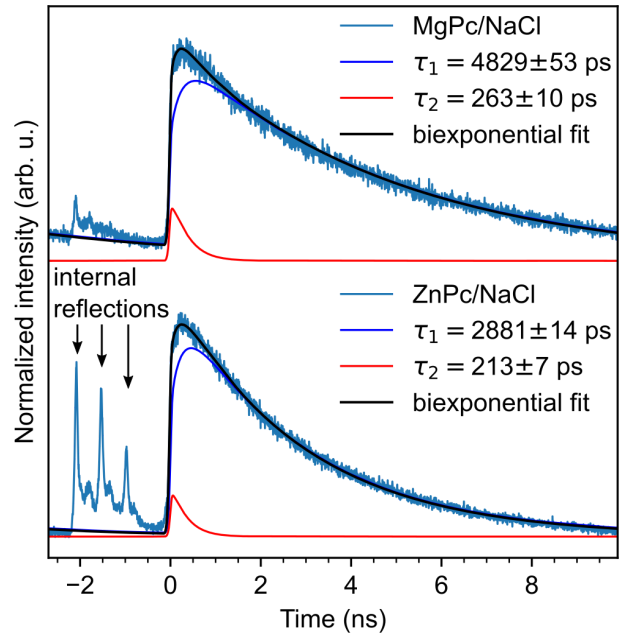

*Fig.3: Biexponential fit of the µPL photon-arrival histograms with a retracted tip for (top) the MgPc total laser power 280 µW, averaging time 180 s, (bottom) the ZnPc, total laser power 600 µW, averaging time 900 s. The zero time is set to the onsets of the photoluminescence from the system.*

The time traces measured in the *µ*PL regime for both considered chromophores manifest a long fluorescence decay spanning a few nanoseconds (Fig.3). Fitting of the acquired time histograms was done with a sum of two exponential functions with different decay rates, convolved with our instrument-response function (IRF), and achieved a good agreement ($R^2 > 0.99$). The IRF was obtained on the clean Ag(111) areas close to the measured molecules by measuring the decay of the nanocavity plasmon (Fig.4a), which is considered to be significantly faster than the molecular emission; the typical electronic dephasing lifetimes are < 100 fs.[10] The need for fitting two decay rates ($\tau_1$ and $\tau_2$) suggests that two different processes are contributing to the total fluorescence yield. The partial yield corresponding to the longer-lived process, responsible for the majority of the signal, is attributed to the molecules adsorbed on top of the 3 ML NaCl, because of their better decoupling from the metal substrate. The lifetimes of 3 ns for ZnPc and 5 ns for MgPc obtained from the fits are close to the values reported for molecules in solutions[30] and are one order of magnitude longer than self-decoupled tetrapodal perylene on Au(111), measured in a similar manner.[31] On the other hand, the contribution with a shorter radiative lifetime is possibly associated with molecules on 2 ML NaCl, where a stronger non-radiative quenching is expected to limit the lifetime.[21] This notion is supported by its much weaker contribution to the total emission in comparison with the longer-lifetime yield (the ratio is 1:37 for ZnPc and 1:36 for MgPc).

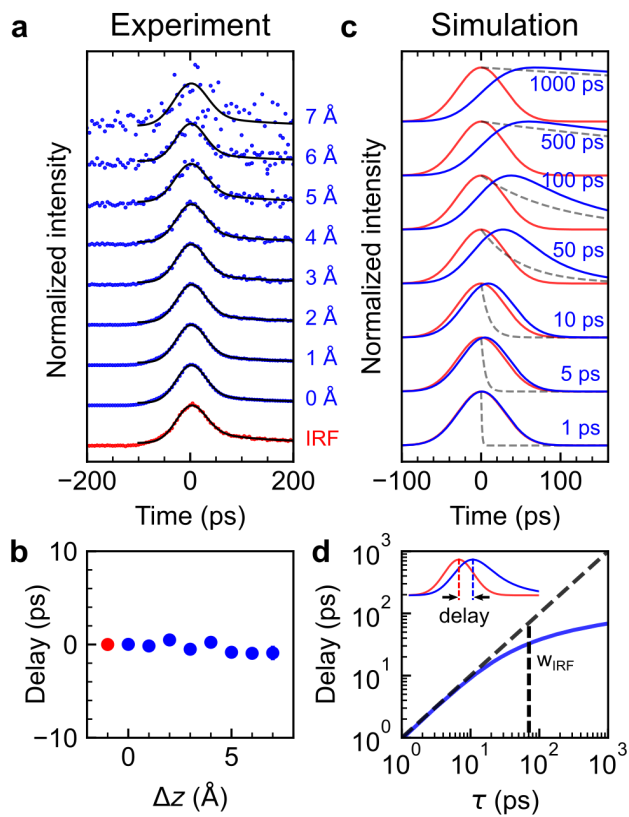

*Fig.4: a) detail of Δz-dependence from Fig.2 after subtraction of the µPL background measured at Δz = 10 Å. The IRF (red) fitted with a modified gaussian (black). b) Delay between the mean arrival times of the IRF and the molecular luminescence ($µ_{mol} - µ_{IRF}$) determined from the fitting in function of the Δz. No measurable change is observed in the given range of. Δz c) Reference simulation of the fluorescence decay histograms with a given Gaussian IRF and its full width at half maximum ($w_{IRF}$ = 70 ps), convolved with an exponential function of variable decay rate. d) The simulated decay-rate-dependent position of the fluorescence decay histogram maximum. Note that for decay rates well below the $w_{IRF}$, the shift of the maximum of the convolution (delay) is asymptotically close to the value of τ.*

For obtaining the excitation lifetimes of molecules located in the nanocavity from the TEPL measurements, the fitting of the exponential decays used for µPL is no longer a valid approach, because the shape of the response and its exponential tail are nearly coincident with the IRF. Therefore, we have adopted an alternative method which allows us to determine the upper limit of the lifetimes with surprisingly higher accuracy. First, the instrumental response function has been fitted with a modified Gaussian function (see Fig. 4a) to extract the mean arrival time of the IRF ($µ_{IRF}$), which is the interval since the laser synchronization pulse defining zero reference time of all measurements in the set, as well as its full width at half maximum ($w_{IRF}$) of 70 ps, stemming from the combined effect of the laser pulse width, jitter of the detector, and the optical pathway. Subsequently, fitting of the Δz - dependent responses of MgPc molecule in the nanocavity shows no significant delay of the peak position ($µ_{mol}$) relative to the IRF (see Fig.4b), proving that the lifetimes are limited to the precision of the measurement and fitting method, *i.e.* < 5 ps in the given Δz range. We illustrate the accuracy of this approach by simulating decay curves for a broad range of lifetime constants, considering a realistic IRF, as shown in Fig. 4c,d. It demonstrates that even for a very small lifetime on the order of < 10% $w_{IRF}$, a departure of the peak position from zero is observable and thus can be quantified. The result of the experiment is similar to the value determined by resonant-absorption STM-PL of $H_2Pc$[16] $S_1$-$S_0$ transition which gives a peak width between 0.2 - 0.9 eV corresponding to 0.7 - 3.3 ps lifetime and points towards a strong enhancement effect of the nanocavity.

In conclusion, by gradually changing the regime between the μPL and TEPL and by using direct TCSPC measurements on single molecular emitters placed on decoupling layers, in a tunable plasmonic nanocavity gap in STM, we have verified the strong increase of radiative decay rate and the associated lifetime shortening, yet without any signature of strong coupling regime. We have determined the upper limit of the excitation lifetime in the STM nanocavity for the MgPc and ZnPc species on 3 ML NaCl. In addition, in the μPL regime, we have found that intrinsic lifetimes of the emitter ensemble on the surface correspond to their lifetimes typically found in solutions. Therefore a pathway is paved toward combined mesoscopic and nanoscopic measurements of excitons of molecular chromophores, their aggregates, and other quantum emitters. We envisage future experiments to determine the lifetimes with a greater precision using the pump-and-probe schemes or GHz-modulation of the incident light.

Methods:
Sample preparation and optical setup
The NaCl was evaporated at 610°C on a clean Ag(111) surface by thermal deposition. ZnPc or MgPc molecules were evaporated at 382°C on the sample held at 10 K, achieving a concentration of around 20 molecules per 100 x 100 $nm^2$. The excitation source in the photoluminescence and the TCSPC measurements was a pulsed supercontinuum laser (Fianium SuperK) with a spectral filtering module (Varia SuperK). Its center wavelength was set to 633 nm, with 10 nm bandwidth. The laser output beam, from an endlessly multimode fiber was collimated by a lens and filtered again using a 633/25 nm bandpass filter, sent through a rotatable half-wave plate and polarized to the plane intersecting the axis of the SPM tip. The beam was reflected into the UHV system by Semrock 633 nm RazorEdge Dichroic™ longpass beamsplitter and focused by an internal lens of the Createc 4K SPM instrument into the tunneling junction area. The final focused laser spot had an approximate diameter of 10 μm. The photoluminescence spectra were measured by an Andor Kymera 328i spectrograph with a 1200 grooves/mm, 500 nm blaze grating.
For the TCSPC measurements, photon arrival time detection was performed using one MPD PDM Series-100 single-photon avalanche detector with 35 ps jitter, together with the Swabian Instruments Time Tagger Ultra pulse counter. The bin size was set to 4 ps. The photon-arrival histograms were accumulated for 90, 360 and 900 s in the measurements of ZnPc near-field, nanocavity plasmons and ZnPc far-field, respectively. The laser power was 100 μW for near-field measurements to avoid light-induced molecular manipulation and 280-600 μW for the far-field measurements.

Fitting procedure
For all the μPL and TEPL TCSPC measurements a corresponding IRF was determined by measuring a plasmonic response of the surface with the tip in the tunneling regime and subtracting the μPL background with the tip retracted by 1 nm. Fitting of the μPL data was done with a biexponential function $A_1 \exp(-t/\tau_1) + A_2 \exp(-t/\tau_2)$ convolved with the IRF. The ratio of the fitting components is calculated as $(A_1\tau_1)/(A_2\tau_2)$. The mean time ($\mu_{IRF}$) and the full width at half maximum ($w_{IRF}$) of the IRF in Fig. 4 were obtained by fitting a Gaussian function, with an additive background accounting for the nonlinear temporal response of the detector: ($A_1 \exp(-2.773(t-\mu)^2/w^2) + A_2 \exp(-2.773(t-\mu)^2/w^2 * \exp(-t/\tau_{IRF}))$) The TEPL data from MgPc were fitted with the same function with fixed $\tau_{IRF}$=108 ps and $A_2/A_1$ = 0.69. Python package LMfit was used for the data fitting.

TEPL tuning

For TEPL measurements, we used tips made of 25 μm diameter Ag wire sharpened by head-on sputtering with a focused ion beam. Further cleaning by Ar$^+$ sputtering was done before insertion into the microscope. The nanocavity plasmon resonance was tuned by voltage pulses and controlled tip indentations into the Ag(111). An effective coupling of the incident laser beam to the nanocavity is typically evidenced by a broad feature in the emission spectrum[29] or as a rigid shift of the field emission resonances, corresponding to the energy of the excitation source.[32]

Acknowledgements: We are thankful to the service provided by Dr. J. Kopeček by sharpening the Ag tips with FIB at the Institute of Physics, Czech Academy of Sciences. We acknowledge the funding from the Czech Science Foundation grant 22-18718S and the support from the CzechNanoLab Research Infrastructure supported by MEYS CR (LM2023051).

Author contributions:
M.Š. and J.D. conceived the experiment. J.D., R.C.C.F and A.S. performed the experiments. J.D. and M.Š. analyzed the data and created the figures. All authors discussed the data and contributed to the writing of the manuscript.